\documentstyle[aps,floats,epsf,twocolumn]{revtex}
\tighten
\begin{document}
\draft

\title{\bf Magnetic Excitations in the Spin-Peierls System
{\mbox{\boldmath $CuGeO_3$}}}

\author{G. Bouzerar, A.P. Kampf, and F. Sch\"onfeld}

\address{ Institut f\"ur Theoretische Physik, Universit\"at zu K\"oln,\\  
Z\"ulpicher Str. 77, K\"oln 50937, Germany.
}
\address{~
\parbox{14cm}{\rm
\medskip
We have calculated the magnetic excitation spectrum in frustrated and dimerized
spin 1/2 Heisenberg chains for model parameters which describe the
thermodynamics and low frequency spin dynamics in the spin--Peierls system 
$CuGeO_{3}$. As a test the chosen model is found to reproduce the lowest Raman 
excitation energy near $30 cm^{-1}$ in the dimerized phase. We establish the 
elementary triplet and singlet excitation branches below the continuum and 
estimate the interchain coupling effects on the dimerization parameter.
}}
\maketitle


\narrowtext
Among the current experimental and theoretical efforts for understanding the 
magnetic properties of linear chain and ladder materials the recent discovery 
of a spin--Peierls (SP) transition at $T_{SP}=14.3K$ in the inorganic compound
$CuGeO_{3}$ \cite{Hase} has attracted particular attention. This interest is 
partially due to the unique structure and available quality of $CuGeO_3$ 
crystals which allow for very detailed studies of the SP phenomenon as well as 
due to the rich phase diagram which evolves in an applied magnetic field 
\cite{Palme} and with doping by substituting $Zn$ for $Cu$ \cite{Lussier} or 
$Si$ for $Ge$ \cite{Renard}, respectively. The magnetic properties of 
$CuGeO_{3}$ arise from $Cu^{2+}$
spin 1/2 moments in weakly coupled $CuO_{2}$ chains which dimerize below 
$T_{SP}$, leading to an alternation of the $Cu$--$Cu$ distance along the chain 
\cite{Bray}. This magnetoelastic transition is driven by the magnetic energy 
gain of antiferromagnetically (AF) coupled $Cu^{2+}$ spins which 
overcompensates the lattice deformation energy \cite{Cross}. In the SP ordered 
phase the copper moments form singlet dimers along the chains with an energy 
gap to spin triplet excitations. Experimentally, the SP nature of the 
transition and the spin excitation gap have been firmly established by 
inelastic neutron scattering (INS), susceptibility, $X$--ray, and 
electron--diffraction experiments \cite{Regnaultrev}.
 
The dominant magnetic intrachain coupling between $Cu^{2+}$ moments in 
$CuGeO_3$ arises from superexchange via the bridging oxygens. Although the 
$Cu-O-Cu$ bond angle is near ${90}^\circ$ and therefore, by the 
Goodenough--Kanamori rules expected to be ferromagnetic, side group effects due
to the hybridization of $O$ and $Ge$ orbitals lead to an effective AF exchange 
interaction \cite{Khomskii}. Furthermore, $Cu-O-O-Cu$ exchange paths lead to an
additional sizeable next--nearest neighbor AF exchange coupling which 
frustrates the magnetic interaction between the $Cu^{2+}$ spins. 

Due to the weak interchain coupling, estimated from the dispersion 
of the magnetic excitations perpendicular to the chains to be an order of 
magnitude smaller than the intrachain exchange \cite{NishiRegnault}, a 1D 
approach for modeling the magnetic properties of $CuGeO_3$ serves as an 
appropriate starting point. Based on the above arguments we use the 1D model 
Hamiltonian 
\begin{equation}
H=J\sum_{i}\left([1+\delta(-1)^i]{\bf S}_i\cdot{\bf S}_{i+1}+\alpha{\bf S}_i
\cdot{\bf S}_{i+2}\right)
\end{equation}
where $i$ denotes the sites of a chain with length $L$ and ${\bf S}_i$ are $S=
1/2$ spin operators. $J>0$ is the intrachain exchange coupling and $\alpha$ the
frustration parameter. The dimerization $\delta$ accounts for the alternation 
of the $Cu-Cu$ distance along the chain in the dimerized phase. For $\delta>0$ 
or for $\alpha>\alpha_c$ the singlet groundstate of this model is dimerized and
a spin gap appears in the excitation spectrum. By numerical scaling analysis it
has been shown that for $\delta=0$ the critical frustration for spontaneous 
dimerization is $\alpha_c\approx 0.2412$ \cite{Okamoto,Castilla}. For $\alpha<
\alpha_{c}$ and $\delta=0$ the system is gapless and renormalizes to the 
Heisenberg fixed point. For the special case $2\alpha+\delta=1$ the groundstate
is known exactly to be a product wavefunction of independent singlet dimers 
\cite{MajumdarShastry}. 

Considerably different parameter values in Eq. (1) have so far been used in 
previous work on $CuGeO_3$ \cite{Riera,Castilla}. Therefore, as a prerequisite 
for further $CuGeO_3$ specific analysis of the model Hamiltonian, we first fix 
the three parameters by using thermodynamic and INS data. With the fixed 
parameter set we evaluate the Raman intensity in the dimerized phase and find 
that the energy of the lowest Raman active singlet excitation is smaller than 
twice the triplet excitation gap $\Delta_{01}$. Thus there exist clearly at 
least {\it two} elementary excitation branches -- singlet and triplet, 
respectively -- below the continuum of excitations starting at $2\Delta_{01}$. 
Previously it was indicated that a singlet excitation might result from a bound
state of two triplet excitations \cite{Uhrig,Kuroe}.

In order to fix the parameters $J$ and $\alpha$ we follow the analysis of 
B\"uchner et al. \cite{Buchner} which is based on susceptibility (up to 
$1000K$) $\chi_b(T)$ data for magnetic fields applied along the crystal 
$b$--direction and thermal expansion data in the uniform ($\delta=0$) phase; 
the latter data set contains implicit information about the magnetic 
contribution to the specific heat. They convincingly conclude that clearly 
$\alpha>\alpha_c$ and from the experimental data for $T>T_{SP}$ the exchange 
couplings are estimated to be near $J=160K$ and $\alpha=0.35$. Indeed, we have 
verified with Lanczos diagonalization techniques applied to Eq.(1) for $\delta=
0$ that with these parameters an extraordinarily good fit to $\chi_b(T)$ in 
the whole temperature range $50K<T<1000K$ can be obtained using the $g$ factor 
$g_b=2.26$ as measured by ESR \cite{Imagawa}. Note that this result for 
$\alpha$ is not in conflict with the work of Castilla et al. 
\cite{Castilla} which favours $\alpha<\alpha_c$. Their arguments were 
based on early neutron scattering data by Harris et al. \cite{Harris} 
which were recently revised by Martin et al. \cite{Martin}.

\begin{figure}
\epsfxsize=8.0cm
\epsffile{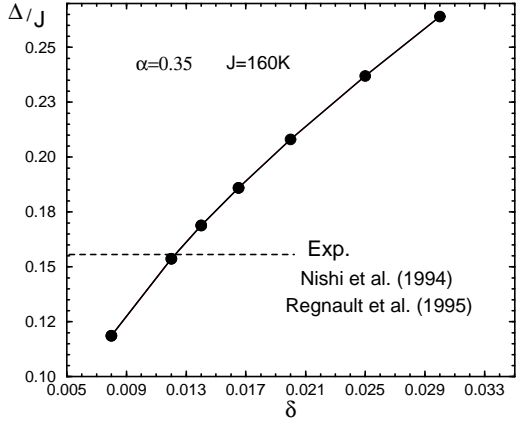}
\caption[]{
Extrapolated singlet--triplet gap vs. dimerization $\delta$ for $\alpha=0.35$.
The dashed line indicates the experimental results from \cite{NishiRegnault}.}
\label{fig1}
\end{figure} 

Given $J$ and $\alpha$ we fix the dimerization parameter $\delta$ from the 
requirement that the excitation spectrum of Eq.(1) reproduces the low 
temperature value of the singlet--triplet gap $\Delta _{01}^{exp}\approx 2.15 
meV=0.156J$ as measured by INS \cite{NishiRegnault}. We have calculated 
$\Delta_{01}$ on finite chains with $L\le 24$ sites as a function of $\delta$ 
using a Lanczos algorithm. In order to extrapolate to the infinite chain limit 
$L\rightarrow\infty$ we use the scaling ansatz
\begin{equation}
\Delta_{01}(L,\delta)=\Delta_{01}(\delta)+\frac{A}{L}\exp{\left(-\frac{L}{L_{01
}(\delta)}\right)}
\label{fiteq}
\end{equation}
and an analogous ansatz for the groundstate energy per site $E_0/L={\it const.
}+(B/L^2)\exp{(-L/\xi)}$. The exponential in each case reflects the gapped 
situation. For the groundstate energy the length scale $\xi$ is the spin--spin 
correlation length as we have verified by a direct calculation of the 
correlation function $\langle S^z(r)S^z(0)\rangle$. The ansatz Eq.(\ref{fiteq}) has been previously motivated and used succesfully by Barnes et al. \cite{Barnes} to study the magnetic excitations in gapped spin ladder systems. In Fig.\ref{fig1} we have plotted the extrapolated results of $\Delta_{01}$ vs. dimerization $\delta$ 
which follow the relation $\Delta_{01}=c_1+c_2\delta^a$ with the Cross--Fisher
exponent $a\approx 2/3$ as we have verified by density matrix renormalization 
group (DMRG) calculations. We observe that the experimental singlet--triplet 
gap value is obtained for $\delta\approx 0.012$ for which the ground state 
correlation length is $\xi\sim 7.3$ lattice spacings.

Having fixed the model parameters we now explore their consequence for the 
Raman intensity. Raman light scattering measures the singlet excitations and is
thus complementary to INS from which the dispersion of the triplet excitation 
has been obtained. Kuroe et al. \cite{Kuroe} and van Loosdrecht et al. 
\cite{Loosdrecht} have observed additional peaks in the Raman intensity which 
appear on cooling below $T_{SP}$. The lowest Raman excitation in the dimerized 
phase is observed at $30 cm^{-1}$, i.e. slightly below $2\Delta_{01}$. 

Within the standard Loudon--Fleury theory of magnetic Raman scattering we write
the relevant part of the Raman operator for the dimerized phase as 
\begin{equation}
H_{R}=\Lambda\sum_i\left({\bf S}_i\cdot{\bf S}_{i+1}+\gamma{\bf S}_i\cdot
{\bf S}_{i+2}\right)
\label{hr}
\end{equation}
for a scattering geometry in which the incoming and scattered photons are 
polarized along the chain \cite{Singh}; experimentally, essentially no 
scattering is observed in other geometries. $\Lambda$ is an overall coupling 
constant while the term proportional to $\gamma$ arises as a consequence of 
frustration \cite{Muthukumar,Singh}. The value of $\gamma$ depends on 
microscopic details. However, it is expected close to the value of $\alpha$ due
to their common microscopic origins. With $H_R$ in Eq.(\ref{hr}) the scattering
intensity is calculated from 
\begin{equation}
I(\omega)\propto -{\rm Im}\langle 0|H_R{1\over\omega+{\rm i}0^+-H+E_0}H_R|0
\rangle \,\, .
\label{ramint}
\end{equation}

\begin{figure}
\epsfxsize=8.0cm
\epsffile{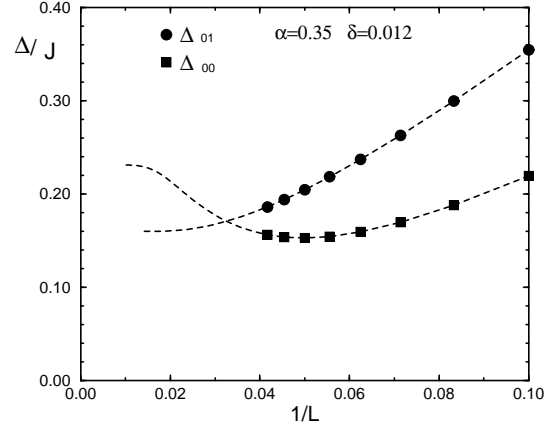}
\caption[]{
Singlet--triplet $\Delta_{01}$ and singlet--singlet gap $\Delta_{00}$ as a 
function of $1/L$. The symbols are exact Lanczos and the dashed lines DMRG 
results.}
\label{fig2}
\end{figure} 

We focus our attention on the lowest energy Raman excitation in $I(\omega)$. In
Fig.\ref{fig2} we have plotted $\Delta_{01}$ and the singlet--singlet gap 
$\Delta_{00}$ as a function of $1/L$ for our fixed parameter set. We present 
both, exact Lanczos diagonalization results for chains up to 24 sites and 
DMRG data for $L\le 100$. For the DMRG calculations we have used the infinite 
system method with periodic boundary conditions keeping 120 states for one 
target state \cite{WhiteChitra}. 

Fig.\ref{fig2} shows that $\Delta_{00}$ has an unusual nonmonotonic scaling 
behaviour. As a finite size effect for short chains, the lowest singlet appears
below the triplet excitation; this has also been noted by Riera et al. 
\cite{Rierabis}. Our DMRG and Lanczos data are in excellent agreement for $L
\le 24$ and accurately confirm the ansatz Eq.(\ref{fiteq}). Without any further
scaling fit the DMRG data allow us to read off directly the $L\rightarrow
\infty$ value of $\Delta_{00}\approx 0.232J$. Obviously we have $\Delta_{00}
\le 2\Delta_{01}$ and $\Delta_{00}$ is close to but $\sim 13\%$ below the 
experimental result for the lowest Raman excitation energy $\Delta_{00}^{exp}=
30cm^{-1}=0.268J $.

Yet, the strong magnetoelastic coupling as underlined by the large 
magnetostriction observed in $CuGeO_3$ \cite{BerndPRL} implies that magnetic 
excitation energies will be renormalized due to the coupling to the lattice and
a similar and presumably larger renormalization is expected from interchain 
coupling. Hence, perfect agreement with the experimental data for both the 
triplet {\it and} the singlet excitation gap can not be expected and we 
consider the above result for $\Delta_{00}$ convincing evidence for a 
consistent parameter choice for $J$, $\alpha$, and $\delta$. We note that the 
contrary conclusions of \cite{Gros} are not based on a proper finite size 
scaling analysis as in Eq.\ref{fiteq}.

\begin{figure}
\epsfxsize=8.0cm
\epsffile{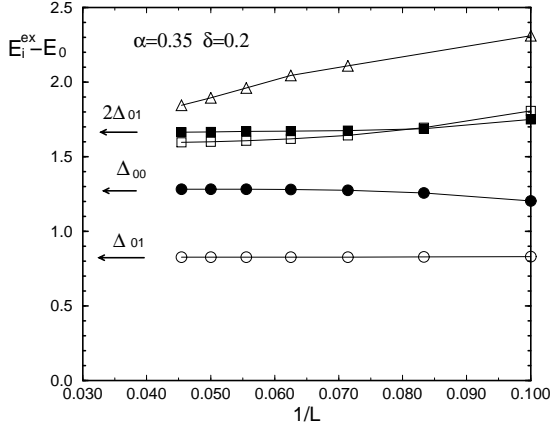}
\caption[]{
Scaling of the low excitation energies (in units of $J$) in the $k=\pi$ momentum subspace. The open (filled) symbols are the triplet (singlet) excitations.}
\label{fig3}
\end{figure} 

In order to analyze the Lanczos excitation spectrum in more detail we consider 
the strong dimerization case $\delta=0.2$ for $\alpha=0.35$ because in this 
regime the finite size effects are very small while we have verified that the 
physics remains essentially unchanged. In fact, the ratio $R=\Delta_{00}/
\Delta_{01}$ changes weakly as we increase the dimerization parameter 
$\delta$, i.e. $R=1.55$ for $\delta=0.2$ to $R=1.49$ for $\delta=0.012$. We 
have plotted in Fig.\ref{fig3} the scaling of the low energy triplet and 
singlet excitations in the subspace of momentum $k=\pi$. We indicate 
$2\Delta_{01}$ in Fig.\ref{fig3} for which we expect to observe the low energy 
edge for the continuum. Surprisingly we observe that also the second triplet 
excitation energy scales below $2\Delta_{01}$. 

\begin{figure}
\epsfxsize=10.0cm
\epsffile{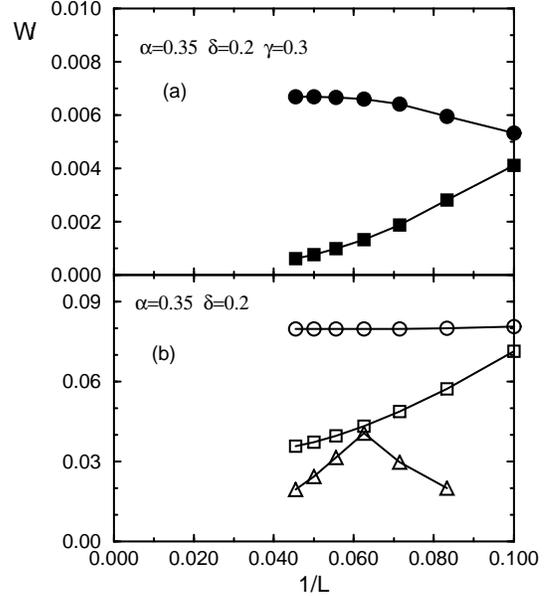}
\caption[]{
(a) Scaling of the weight of the 2 lowest Raman singlet excitations calculated 
with $\gamma=0.3$ in Eq.\ref{hr}.
(b) Spectral weight of the low energy triplet excitations in $S(\pi,\omega)$ 
vs. $1/L$. For the lowest excitation the weight is multiplied by $0.2$.
In (a) and (b) the same 
symbols are used as in Fig.\ref{fig3}.}
\label{fig4ab}
\end{figure} 

Complementarily we calculate the spectral weight of these excitations in the 
dynamical structure factor $S(\pi,\omega)$ and the Raman intensity $I(\omega)$,
respectively, as given by $W_{i}=|\langle i|\hat{O}|0\rangle|^2$ where $|0
\rangle $ is the groundstate and $|i\rangle$ is an excited state with momentum 
$k=\pi$. The operator $\hat{O}$ is $H_{R}/\sqrt{L}$ for the singlet and $S^z(k)
=(1/\sqrt{L})\sum_l\exp{({\rm i}k\,l)}S^{z}_l$ for the triplet excitations. In 
Fig.\ref{fig4ab}b we show results for $W_i$ of the 3 lowest triplet excitations
as a function of $1/L$. It appears clearly that $W_i$ for the two lowest 
triplets scales to a finite value, identifying them as elementary excitations 
below the continuum. Contrary, $W_i$ for the third excitation in the triplet 
subspace scales to 0 while its energy extrapolates to $2\Delta_{01}$, i.e. the 
onset of the continuum. This is also reflected by the kink in the scaling data 
for $L=16$. In Fig.\ref{fig4ab}a we perform the same scaling of $W_i$ for the 
singlet excitations in the Raman intensity evaluated for $\gamma=0.3$ in 
Eq.(\ref{hr}). The conclusions remain unchanged if $\gamma$ is varied. 
$W_i$ of the lowest singlet excitation obviously scales to a finite 
value while the weight of the next higher energy singlet excitation 
scales to 0.

\begin{figure}
\epsfxsize=8.0cm
\epsffile{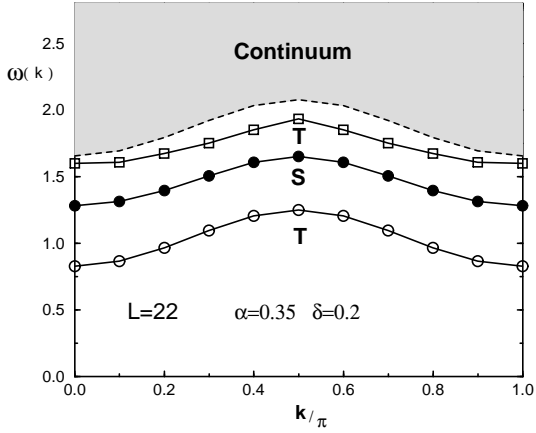}
\caption[]{
Dispersion $\omega (k)$ (in units of $J$) of the elementary triplet and singlet excitations. The continuum from all spin sectors above $2\Delta_{01}(k)$ is indicated by the shaded area.}
\label{fig5}
\end{figure}

From this finite size scaling analysis we conclude that there are 3 elementary 
excitations, 2 triplet and 1 singlet, for $\delta=0.2$; the two lower 
excitations clearly persist when $\delta$ is decreased to $0.012$ -- the model 
parameter value relevant for $CuGeO_3$. The singlet excitation at $k=0$ (or 
$k=\pi$) is identified with the $30cm^{-1}$ peak in the Raman intensity while 
the lower triplet corresponds to the spin excitations measured in INS. For the 
existence of the second triplet near the edge of the continuum, however, a 
definite conclusion could not be reached for $\delta=0.012$. The results for 
$\delta=0.2$ are summarized in Fig.\ref{fig5} which shows the dispersions for 
the 3 elementary excitations.

So far we have used a strictly 1D approach to describe the magnetic excitations
in the SP phase of $CuGeO_3$. Yet, for a more realistic model the interchain 
coupling via the oxygen atoms that two adjacent $CuO_2$ chains have in common 
cannot be ignored.
Indeed, the $b$--direction dispersion of the triplet excitation measured in INS
is not small and its energy varies from its maximum value of $5.8meV$ to 
$2.1meV$ at the Brillouin zone (BZ) center \cite{NishiRegnault}. In a first 
attempt to include interchain coupling effects into an effective 1D spin 
Hamiltonian we may determine the dimerization parameter from the BZ averaged 
triplet excitation gap $\approx 3.95meV$ rather than from its BZ center value 
used above. Keeping $J$ and $\alpha$ fixed as before, because they are 
determined from data in the uniform phase, then leads to an estimate $\delta
\sim 0.036$, i.e. an increase by a factor of $3$. 
Assuming a similar $b$--direction
dispersion also for the singlet excitation branch we find that the BZ center 
value of the singlet gap remains almost unchanged and thus our conclusions are
still valid. 

In conclusion, we have fixed the parameters of a frustrated and dimerized 
Heisenberg chain model to describe the thermodynamics and the low energy 
magnetic excitations at low temperatures in $CuGeO_3$. The internal consistency
of the parameter choice is verified by almost quantitatively reproducing 
the energy of the lowest singlet excitation in the dimerized phase 
as measured by Raman light scattering. Furthermore, we have shown 
the existence of at least 1 elementary triplet and 1 single excitation 
branch below the continuum. While the effects 
of weak interchain coupling on the spin dynamics still remain to be explored 
the $J-\alpha-\delta$ model Hamiltonian with fixed parameters will serve as a 
basis for studying doping and magnetic field dependent phenomena in $CuGeO_3$. 

We thank E. M\"uller--Hartmann, B. B\"uchner, G. Uhrig and K. Hallberg for 
instructive discussions. We are particularly grateful to B. B\"uchner for 
informing us about unpublished susceptibility data and results for the 
determination of $J$ and $\alpha$ in the uniform phase. Research was 
performed within the program of the 
Sonderforschungs\-be\-reich 341 supported by the Deutsche 
Forschungsgemeinschaft (DFG). A.P.K. gratefully acknowledges support through a 
Heisenberg fellowship of the DFG.

\end{document}